# Extraire et réutiliser des patrons de conception à partir de Learning Games existants

Iza Marfisi-Schottman, Claudine Piau-Toffolon
LUNAM Université, Université du Maine, EA 4023, LIUM, 72085 Le Mans, France
iza.marfisi@univ-lemans.fr, claudine.piau-toffolon@univ-lemans.fr

**Résumé.** Les Learning Games (LGs) sont des outils pédagogiques très prometteurs, mais leur conception reste encore expérimentale. Inspirés par les méthodes de conception à base de patrons, préconisées dans le domaine de l'éducation, nous proposons une méthodologie et un modèle capable d'analyser le scénario de LGs éprouvés sur le terrain, afin d'en extraire des patrons de conception. La méthodologie et le modèle proposés ont permis d'extraire neuf patrons de conception à partir de l'analyse de deux LGs utilisés depuis une dizaine d'années. Ces patrons se sont avérés très utiles puisque plus de la moitié ont été adoptés par des équipes de concepteurs, lors de la création de LGs pour des contextes d'utilisation similaires aux deux LGs existants.

**Mots-clés.** Jeux sérieux, learning games, patrons de conception, modèle de scénario

**Abstract.** Learning Games (LGs) are promising pedagogical tools but their design still remains experimental. Inspired by design-pattern based methods, recommended in educational domains, we propose a methodology and a model to analyze the scenario of LGs, which have proven to be effective, in order to extract design patterns. The proposed methodology and model allowed us to extract nine design patterns from the analysis of two LGs, which have been actively used in schools for nearly ten years. These design patterns proved to be very useful because half of them where adopted by teams of designers, in the process of creating LGs, for similar contexts to the ones of the existing LGs.

## 1. Introduction

Les Learning Games (LGs) sont des applications informatiques qui utilisent des ressorts ludiques tels que la compétition, les récompenses ou tout simplement la curiosité pour catalyser l'attention des apprenants et faciliter leur apprentissage [1]. De nombreux LGs ont démontré des capacités à impliquer les apprenants dans leurs formations, à faciliter l'acquisition de compétences ou encore à stimuler la mémoire [1]. Cependant, ces qualités sont loin d'être facilement reproductibles lors de la conception d'un nouveau LG. Ceci est principalement dû à la nature pluridisciplinaire des LGs, qui rend leur création très complexe, car elle combine les problématiques liées à la conception pédagogique, celles liées à la conception d'un jeu, mais également le défi supplémentaire de concevoir un jeu qui soit aux services des objectifs pédagogiques [2].

Dans cet article, nous proposons de partir d'un modèle issu de nos travaux précédents sur l'environnement de conception LEGADEE[1] (LEarning GAme DEsign Environment) pour identifier les éléments pédagogiques et la façon dont ils sont instanciés dans un jeu avec des éléments et un scénario ludique. L'idée est d'utiliser ce modèle en *reverse engineering*, pour extraire des caractéristiques récurrentes des LGs existants, et ainsi pouvoir les reformuler en patrons de conception. Après avoir étudié les travaux existants relatifs à la conception des LGs, nous utilisons le modèle proposé pour décrire le scénario de deux LGs, qui sont utilisés depuis une dizaine d'années. À partir de l'analyse de ces deux scénarios et des entretiens avec les enseignants qui les utilisent, nous extrayons ensuite les mécanismes qui semblent essentiels à ces LGs et les reformulons sous la forme de neuf patrons de conception. Dans la quatrième partie, nous montrons ensuite comment ces patrons ont été utilisés pour aider des équipes de concepteurs à créer des LGs dans des contextes similaires aux deux LGs analysés. La démarche adoptée est proche des méthodes « centrées utilisateurs » et des préconisations du « Design Based Research » [Wang et Hannafin 05]. Plusieurs méthodes d'évaluation ont été adoptées dans une démarche itérative où les objets ont évolué en cours de conception. Pour finir, nous proposons des pistes d'outils pour fournir de l'aide adaptée et automatique, afin d'améliorer la qualité d'un LG en cours de conception.

## 2. Conception de LG: Etat de l'art

La conception d'un LG consiste à élaborer une série d'activités pédagogiques, intégrées dans un scénario ludique permettant à l'apprenant d'atteindre un ou plusieurs objectifs d'apprentissage [3]. Pour ce faire, il s'agit de mettre en œuvre un processus de conception impliquant la collaboration d'experts pédagogiques et de *game designer*, des acteurs ayant pourtant des compétences, du vocabulaire des méthodes de travail très différentes [4]. Afin de faciliter cette collaboration, plusieurs chercheurs ont proposé des méthodologies qui identifient clairement les interactions et les tâches de chaque acteur tout au long de la création d'un LG [2]. De façon complémentaire, d'autres études prometteuses portent sur des outils pour faciliter la collaboration et notamment sur des méthodes à base de patrons de conception, un procédé commun aux *game designers* et aux experts pédagogiques.

En effet, bien que chaque *game designer* semble utiliser sa propre technique de créativité, les recueils de patrons de *gameplay* sont largement répandus dans les écoles de *game design* [5]. Plusieurs plateformes participatives, telles que mediaWiki[2], proposent également des centaines de patrons pour favoriser les échanges entre *game designers*. On y trouve par exemple le patron « niveaux finals », qui consiste à structurer le scénario du jeu avec des niveaux ou des cinématiques, qui concluent chaque grande partie de l'histoire, afin que le joueur sente qu'il progresse dans la narration. On y trouve également le patron « statistique de jeu » qui consiste à collecter tous les scores des joueurs dans un tableau de bord, afin de les inciter à progresser, soit pour améliorer leurs propres scores, soit pour battre les autres joueurs.

---

[1] http://www-lium.univ-lemans.fr/legadee/
[2] http://129.16.157.67:1337/mediawiki-1.22.0/index.php/Category:Patterns

Le vocabulaire et le format standardisé dans lequel ces patrons sont écrits (i.e. description, exemples de jeux connus utilisant ce patron et les relations avec d'autres patrons) servent notamment de langage de référence pour faciliter la collaboration entre les membres d'un projet. Les approches de conception à base de patrons sont également préconisées dans le domaine de l'éducation [6][7][8] et ont fait l'objet de diverses propositions d'outils d'édition pour la conception de scénarios pédagogiques [9][10][11][12]. Ces approches facilitent la négociation et la conception participative [6]. Quelques chercheurs ont d'ailleurs repris ces patrons pour les adapter aux objectifs spécifiques des LGs [13][14], mais ces bibliothèques de patrons restent incomplètes, notamment en terme d'exemples. Ceci peut s'expliquer par le fait que leurs patrons sont issus de recherche en *game design* et en éducation et non directement de l'analyse de LGs qui ont fait leurs preuves sur le terrain. Il semble pourtant primordial d'analyser ces LGs, afin d'extraire les caractéristiques récurrentes qui font leurs succès. Cependant, comme nous l'avons montré précédemment, la nature pluridisciplinaire des LGs conduit à des scénarios très complexes, combinant des éléments de nature pédagogique et ludique et il est très difficile de dissocier ces éléments afin d'identifier les mécanismes qui donnent lieu aux synergies recherchées.

## 3. Modèle pour l'analyse des scénarios de Learning Games

Comme nous venons de le souligner, il est particulièrement difficile d'analyser les mécaniques qui sont à l'origine de l'efficacité des LGs parce que leurs scénarios sont composés d'une multitude d'éléments pédagogiques et ludiques entremêlés. Afin de faciliter leur identification et la nature de leurs liens, nous proposons d'utiliser un modèle issu de l'environnement de conception LEGADEE[3]. Cet environnement facilite la conception collaborative de LGs en proposant une méthodologie, un modèle et un ensemble d'outils pour guider l'équipe de concepteurs [2]. La particularité du modèle qu'il intègre, est de dissocier les éléments pédagogiques (*Structure pédagogique, Compétences et Participants de la formation*) et la façon dont ils sont mis en scène dans un contexte attractif (*Scénario ludique, Personnages du jeu*), mais il permet surtout d'identifier la nature des associations entre ces éléments pédagogiques et leurs contextualisations ludiques (liens d'association en gras dans la Fig. 1). Cette particularité, qui n'est pas présente dans d'autres modèles de scénario LGs, semble être une bonne base pour nos besoins d'analyse de LGs existants. Dans la suite, nous présentons donc ce modèle (Fig. 1).

### 3.1. Eléments pédagogiques

La ***Structure pédagogique*** d'un LG est composée en trois niveaux de granularité : les *Modules* correspondent à un découpage similaire au chapitre d'un cours, les *Actes* représentent des sessions d'activités qui ont un sens au niveau pédagogique et les *Activités* de bas niveau comme lire, écrire, évaluer, discuter, dessiner… Chaque *Module*, *Acte* et *Activité* possède également un titre et un objectif pédagogique sous la forme textuelle et doit être relié à au moins une **Compétence**[4]. Ils peuvent également être reliés à un ou plusieurs **Participants** à la formation qui représente un rôle (apprenant, tuteur) ou une équipe de personnes. Ce modèle reprend certains concepts définis par le standard IMS-LD[5] comme les *Actes*, *Activités*, et *Participants,* mais propose également le moyen de structurer la formation en *Modules* plus abstraits, ce qui faisait défaut au standard [15]. Dans la suite, nous allons voir comment ces éléments sont instanciés dans le scénario du jeu avec une mise en scène ludique.

---

[3] voir supra

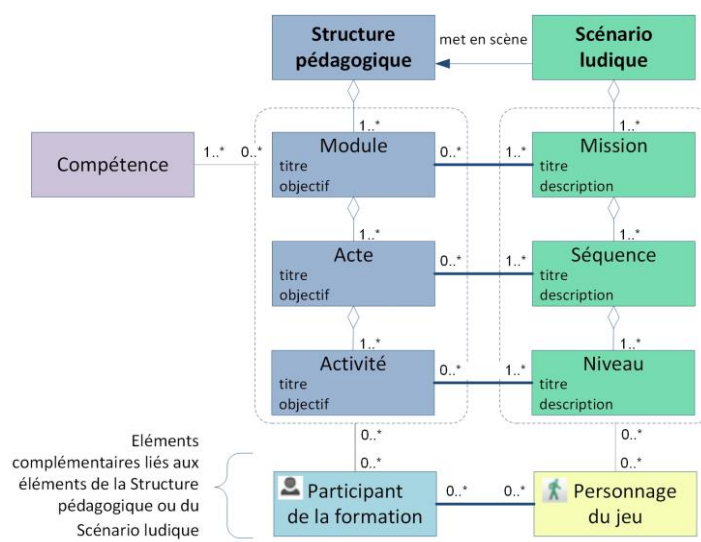

**Fig. 1.** Modèle LEGADEE de scénario de Learning Games

### 3.2. Eléments de mise en scène ludiques

Le *Scénario ludique* est organisé, comme la *Structuration pédagogique*, selon une granularité à trois niveaux. Les **Missions** désignent une grande partie du jeu (ex. introduction à l'intrigue du LG et création de son avatar, déterminer la trajectoire de la fusée en solvant des équations). Les **Séquences** désignent une série de plans qui forment une unité narrative (ex. la découverte de l'entreprise, la recherche d'un indice). Certaines ne sont pas directement liées à l'histoire, par exemple des *Séquences* de test qui visent à valider les connaissances de l'apprenant ou des *Séquences* de briefing et de débriefing pendant lesquelles le formateur peut revenir sur les concepts et faire le parallèle avec les situations réelles. Les **Niveaux**, ou le terme *game level*, couramment employé par les *game designers* [5], peut être apparenté à un plan cinématographique dans le sens où la démarcation entre niveaux peut être liée à un changement de décor (ex. changement du fond, de la couleur), du contenu (ex. nouveaux obstacles, augmentation de la difficulté) ou d'une combinaison des deux. Chaque *Missions*, *Séquences* et *Niveaux* possède également un titre et une description textuelle. De plus, ils peuvent être reliés à des **Personnages** du jeu (ex. héros, méchant).

---

[4] défini comme un « ensemble organisé, cohérent et finalisé de connaissances et comportements s'exerçant dans un contexte et contrôlables par des résultats » [2, p156]
[5] http://www.imsglobal.org/learningdesign

### 3.3. Associations entre les éléments pédagogiques et leur mise en scène ludique

Comme vous pouvez le voir dans la Fig. 1, une *Mission* peut être purement ludique (ex. introduction à l'intrigue du LG et création de son avatar) ou mettre en scène un ou plusieurs *Modules* (ex. la *Mission* où il faut déterminer la trajectoire de la fusée est reliée au *Module* portant sur les équations). De la même façon, une *Séquence* peut mettre en scène zéro, un ou plusieurs *Actes* et un *Niveau* peut mettre en scène zéro, une ou plusieurs *Activités*. De plus, ces liens de « mise en scène » transfèrent les *Compétences* et *Participants* liés aux éléments de la *Structure pédagogique*. Ainsi les éléments du *Scénario ludique* (*Missions*, *Séquences*, *Niveaux*) sont reliés, par capillarité, à des *Compétences* et *Participants*. Pour finir, un *Personnage* du jeu peut aussi être relié à zéro, un ou plusieurs *Participants* de la formation (ex. les apprenants jouent souvent le rôle du héros de l'histoire).

Maintenant que nous avons décrit le modèle, nous allons l'utiliser pour analyser le scénario de LGs existant.

## 4. Extraction de patrons à partir de Learning Games existants

Comme nous l'avons montré dans la section précédente, la structure du modèle proposé nous semble pertinente pour analyser les scénarios de LGs existants, afin de trouver des éléments ou des assemblages d'éléments récurrents, pour définir des patrons de conception. Cependant, même si nous nous limitons aux LGs qui ont fait la preuve de leur efficacité, ils représentent des typologies de LGs très variés, vis-à-vis des compétences qu'ils font acquérir et de leurs contextes d'utilisation (i.e. profil des apprenants, LG à utiliser en classe ou seul à la maison…). Dans un premier temps, nous proposons donc de restreindre notre analyse à quelques LGs du même type. En effet, il est beaucoup plus probable de trouver des invariants, et de pouvoir ainsi tester l'efficacité de notre modèle, en analysant une sous-catégorie de LGs.

Dans cette perspective, nous nous sommes joints au projet ANR JEN.lab[6], dont le but est de proposer des modèles pour concevoir des LGs bien particuliers : les Jeux Epistémiques Numériques (JENs). Cette sous-catégorie de LGs a la particularité de proposer des problèmes complexes, pluridisciplinaires et non déterministes et de placer les apprenants dans un contexte réaliste dans lequel ils doivent mener leurs activités avec le savoir-faire et les méthodes de vrais professionnels [16]. Ces jeux sont également conçus pour être utilisés en classe, avec l'enseignant. De plus, grâce aux partenaires du projet JEN.lab, nous avons pu avoir accès à deux JENs éprouvés.

---

[6] http://eductice.ens-lyon.fr/EducTice/recherche/jeux/jen-lab/resume-jenlab

Dans la suite, nous présentons comment nous avons modélisé et analysé le scénario de ces deux JENs avec le modèle LEGADEE et explicitons les neuf patrons de conception que nous avons réussi à extraire à partir de ces deux scénarios.

### 4.1. Modélisation et analyse de deux Jeux Epistémiques Numériques

Avant de décrire leur modélisation, nous présentons succinctement les deux JENs que nous avons sélectionnés pour leurs qualités pédagogiques et ludiques, éprouvés depuis une dizaine d'années, par des enseignants :

- Le JEN *Land Science*[7] (LS) (environ 10h de jeu) place les apprenants dans le rôle de jeunes stagiaires embauchés par une entreprise d'urbanisme, afin de repenser la composition d'un quartier central de leur ville. Le but pédagogique du jeu est d'entraîner les apprenants à adopter une stratégie d'expert urbaniste pour résoudre ce problème complexe : prendre en compte tous les avis des habitants de la ville, utiliser les outils adéquats pour prendre une décision et être capables de présenter et justifier ses choix. L'utilisation de ce jeu par plus de 40 lycéens montre qu'il favorise l'acquisition de compétences portant sur l'urbanisme et l'écologie ainsi que des valeurs civiques [17].
- Le JEN *Puissance7*[8] (PU) (environ 12h de jeu, soit 3 séances de 4h) place les apprenants dans la position d'une équipe de consultants, embauchés par une entreprise de distribution qui rencontre des difficultés à livrer ses clients. Le jeu vise à faire découvrir aux étudiants les grandes phases de la méthode de résolution de problèmes et leur faire mettre en œuvre un certain nombre d'outils professionnels tel que les diagrammes de Pareto, d'Ishikawa et les techniques de brainstorming. Bien qu'il n'y ait pas d'étude scientifique sur l'efficacité de ce jeu, il présente indéniablement des avantages puisqu'il est utilisé depuis plus de 17 ans (première version en 1998) pour les cours de 3[ème] année aux départements Génie Industriel, Informatique et Technique de Commercialisation de l'école d'ingénieur INSA de Lyon.

Nous avons modélisé le scénario de ces deux JEN en détail avec le modèle LEGADEE[9]. Tout d'abord, il est important de noter que nous n'avons pas rencontré de difficultés à modéliser ces LGs puisque nous avons pu renseigner tous les éléments qui les caractérisent, c'est-à-dire les étapes du jeu, leurs enchaînements, les compétences pédagogiques à acquérir, les acteurs (enseignants et élèves), les rôles qu'ils interprétaient ainsi que leurs interactions. Notons, tout de même, que cette modélisation requiert une bonne maîtrise du modèle LEGADEE. De plus, certaines informations, comme la durée d'une étape ou la nature des interactions entre les participants (i.e. compétition, collaboration), sont simplement indiquées sous forme textuelle dans la description des éléments du *Scénario ludique* et non modélisée avec des classes spécifiques. Ceci ne pose pas de problème pour la modélisation ni pour une analyse manuelle des scénarios, mais c'est un point qui pourrait être amélioré

---

[7] http://edgaps.org/gaps/projects/land-science/
[8] http://gi.insa-lyon.fr/files/rte/Puissance7-v2004.pdf
[9] Les modélisations complètes de PU et LS sont accessibles sur http://www-lium.univ-lemans.fr/legadee/ -> Tester Legadee

pour faciliter une analyse automatique de scénario. En outre, la modélisation avec le modèle LEGADEE nous a obligés à comprendre en détail les mécanismes des LGs. Afin de valider la justesse de nos modélisations, nous avons interrogé les enseignants qui les utilisent et nous avons également joué à PU avec un groupe d'étudiants.

L'analyse manuelle des scénarios de LS et PU, afin de trouver des éléments ou des combinaisons d'éléments présents dans les deux JENs, nous a permis d'identifier les sept patrons suivants :
- P1. Teaser de jeu
- P2. Problèmes pluridisciplinaires
- P3. Personnifier des experts
- P4. Explorer différentes solutions
- P5. L'enseignant comme soutien
- P6. Briefing
- P7. Débriefing

Afin de prendre en compte les spécificités de chaque jeu, nous avons également demandé aux enseignants de préciser les mécanismes qui leurs semblaient important. Nous avons ainsi identifié deux autres patrons :
- P8. Travailler en équipe multi point de vue (trouvé dans le scénario de LS)
- P9. Rapport d'analyse post-jeu (trouvé dans le scénario de PU)

Afin d'illustrer ces patrons de conception, nous décrivons les patrons P1 et P5 dans la partie suivante. Les autres patrons sont décrits en annexe.

### 4.2. Exemples de patrons de conceptions extraits

Afin de faciliter la lecture, nous présentons les patrons dans un format classique de patron de conception [5] : une définition, des exemples concrets de l'implémentation du patron dans LS et PU et une représentation visuelle avec le modèle LEGADEE.

**P1. Teaser de jeu**
***Définition*** : le LG commence avec un *teaser*, sous la forme d'une vidéo, d'un email ou d'une présentation orale, dans lequel la mission principale du LG est expliquée. Il doit également mettre les apprenants et l'enseignant dans le rôle de leurs personnages.
***Exemples*** :
- LS : au début du jeu, les apprenants reçoivent un email de Maggy, la chef d'une entreprise d'urbanisme, qui leur annonce qu'ils sont embauchés comme stagiaires et qu'ils ont pour but de repenser l'organisation d'un quartier du centre-ville. L'enseignant, qui joue le rôle d'un collègue serviable, leur donne plus de détail sur leur mission et les aide à s'organiser (Fig. 2, gauche).
- PU: au début du jeu, l'enseignant constitue des groupes de 3 élèves et annonce qu'ils viennent d'être embauchés, en tant que consultants, par une entreprise pour trouver une solution à leurs difficultés de livraisons (Fig. 2, droite).

***Représentation avec le modèle LEGADEE :*** la première *Mission* du scénario est une introduction au jeu. Elle est reliée aux *Personnages* principaux, joués par les apprenants et l'enseignant.

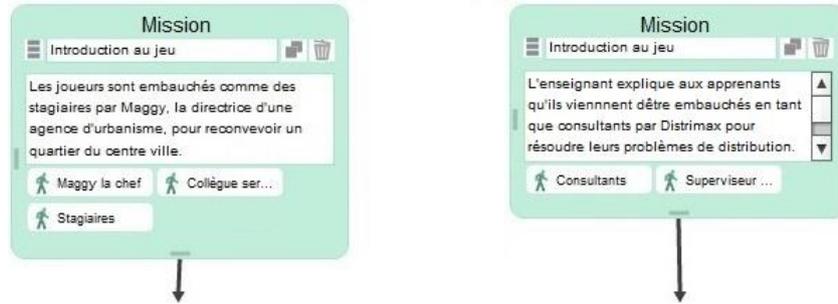

**Fig. 2.** Modélisation de P1. Teaser de jeu (LS-gauche, PU-droite)

### P5. L'enseignant comme soutien
*Définition* : l'enseignant tient le rôle de conseiller bienveillant. Afin de maintenir cette relation de confiance, les évaluations doivent être faites après le jeu (voir P9) ou pendant le jeu, mais de façon dissimulée, en contrôlant secrètement un autre personnage.
*Exemples :*
- LS : l'enseignant principal et les tuteurs pédagogiques qui participent au LG jouent le rôle de collègues serviables nommés Kurt, Sydney et Kira, qui guident chacun un groupe de stagiaires tout au long du jeu et fournissent des conseils. L'enseignant principal joue également secrètement le rôle de Maggy, la chef d'entreprise, qui valide le travail des stagiaires. Ce rôle n'est pas connu des apprenants, car les contacts qu'ils ont avec Maggy se font uniquement via email (Fig. 3, gauche).
- PU : l'enseignant joue le rôle de superviseur de l'équipe de consultants qui est là pour les guider et les encourager tout au long du jeu. À aucun moment, il ne juge ou évalue les performances des apprenants pendant le jeu. L'évaluation se fait sur le rapport d'analyse, rédigée une semaine après le LG (voir P9) (Fig. 3, droite).

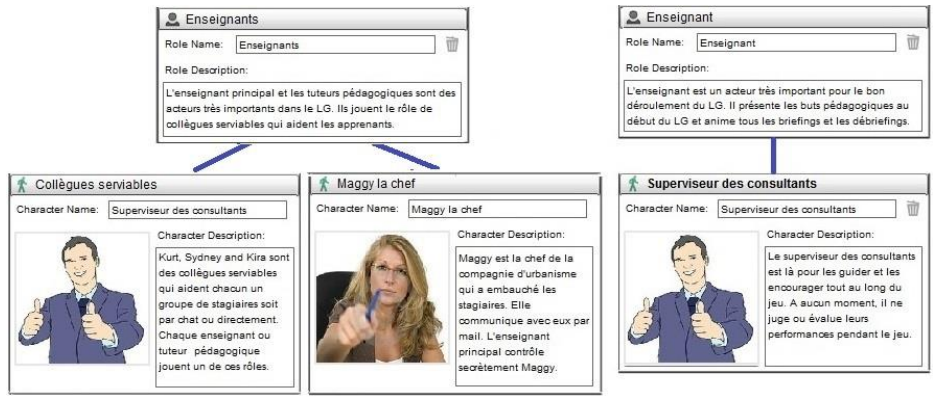

**Fig. 3.** Modélisation de P5. L'enseignant comme soutien (LS-gauche, PU-droite)

***Représentation avec le modèle LEGADEE :*** le ou les enseignants sont modélisés par un *Participant* qui est relié à au moins un *Personnage* du jeu, qui a pour mission d'aider les personnages joués par les apprenants. Il peut également être relié à un *Personnage* dont le but est d'évaluer les apprenants, mais ce rôle ne doit pas être connu des apprenants.

Dans la suite de cet article, nous allons voir comment ces patrons de conception peuvent être utilisés pour guider la conception de nouveaux LGs similaires.

## 5. Aide à la conception

Comme nous l'avons vu dans l'introduction, la conception de LG reste expérimentale et leur qualité souvent décevante. Nous avons donc voulu analyser des LGs efficaces existants pour en extraire des patrons de conception. Dans cette partie, nous allons décrire comment nous avons utilisé les patrons issus des deux JENs analysés et le modèle LEGADEE pour aider quatre équipes de concepteurs à améliorer leurs JENs en cours de conception. Bien que leurs profils aient été variés (i.e. 3 enseignants, 1 chercheur et 11 étudiants de niveau L3), tous les concepteurs étaient débutants en création de LGs.

Tout d'abord, nous avons présenté le modèle LEGADEE aux concepteurs afin qu'ils puissent concevoir leurs LGs avec. Une fois que la trame générale de leur LG était finie (modélisation des *Modules*, *Actes*, *Missions* et *Séquences* terminés), ce qui a pris plusieurs semaines pour certaines équipes, nous leur avons présenté les neufs patrons de conception ci-dessus en les illustrant avec les scénarios de LS et PU. Les concepteurs avaient également accès à la modélisation complète des deux JENs. Afin de valider si les patrons leurs avaient été utiles, nous avons ensuite analysé leurs LGs pour voir s'ils avaient intégré certains des patrons après la présentation. Nous avons également recueilli leurs avis avec des interviews semi-dirigés pour définir les perspectives d'amélioration.

### 5.1. Intégration des patrons de conception

Comme l'indique le Tableau 1, les concepteurs avaient déjà intégré certains des patrons naturellement, avant que nous ne leurs présentions. Leurs LGs proposaient tous des problèmes pluridisciplinaires (P2), deux d'entre eux intégraient des teasers de jeu (P1) et deux autres plaçaient les apprenants dans le rôle d'experts (P3).

**Tableau 1.** Intégration de patrons de conception avant et après leur présentation

| Équipe de concepteur | Patrons déjà intégrés | Patrons ajoutés |
|---|---|---|
| LG1 (3 concepteurs) | P2, P3 | P1, P5, P7 |
| LG2 (3 concepteurs) | P1, P2, P3 | P4 |
| LG3 (4 concepteurs) | P1, P2 | P3, P5, P4 |
| LG4 (4 concepteurs) | P2, P4, P6 | P1, P3, P5, P7, P8 |

Nous pouvons également constater que la présentation des patrons de conception était utile puisqu'à la suite de celle-ci, toutes les équipes ont ajouté de 1 à 5 patrons supplémentaires. Deux équipes ont ajouté des teasers de jeu (P1) soit sous la forme de cinématographie (LG1), soit sous la forme d'un simple email avec une présentation orale de l'enseignant (LG4). Deux équipes ont également modifié leurs LGs de façon à ce que les apprenants soient clairement mis dans le rôle d'experts (P3), en jouant le rôle de pilotes de fusée (LG3) ou de scientifiques élus pour sauver l'humanité (LG4). Trois des équipes ont également renforcé la place de l'enseignant comme une aide dans le jeu (P5) même si celui-ci n'est pas toujours associé à un personnage du jeu. Parmi ces équipes, deux ont ajouté quelques phases de débriefing menées par l'enseignant (P7). En outre, deux équipes ont modifié le scénario de leurs jeux afin de le rendre moins linéaire (P4). Ainsi, deux *Séquences* du LG2 proposent plusieurs solutions et les joueurs peuvent choisir entre deux stratégies pour gagner le LG3. Pour finir, la notion d'appartenance à des guildes avec des points de vue différents a été rajoutée au LG4 de façon à enrichir les débats entre joueurs (P8).

### 5.2. Analyse de l'interview et du questionnaire

L'effet positif de la présentation des patrons de conception que nous avons observés lors de l'analyse des scénarios est renforcé par les réponses des 14 concepteurs au questionnaire. En effet, à la question « Est-ce que le fait de visualiser le scénario de PU et LS et la présentation sur les patrons de conception vous a aidé à améliorer votre LG ? », 2 ont répondu « beaucoup », 11 « plutôt oui », seulement 1 « plutôt non » et aucun « pas du tout ». Lors des interviews, 8 concepteurs ont dit avoir apprécié le fait de pouvoir visualiser leurs scénarios et celui des exemples de LGs avec le même modèle et 6 ont apprécié les explications orales accompagnant la présentation des patrons.

## 6. Conclusion et perspectives

Dans cet article, nous avons proposé le modèle LEGADEE. Ce modèle de scénario de LGs a la particularité d'identifier la nature des associations entre les éléments pédagogiques du LGs (*Structure pédagogique, Compétences et Participants de la formation*) et leurs contextualisations ludiques (*Scénario ludique, Personnages du jeu*). Ce modèle nous a permis d'analyser et de comparer le scénario de deux Jeux Epistémiques Numériques (JENs) existants (une sous-catégorie de LGs), afin d'en extraire des associations d'éléments récurrents. Cette analyse nous a permis de définir neuf patrons de conception : P1. Teaser de jeu, P2. Problèmes pluridisciplinaires, P3. Personnifier des experts, P4. Explorer différentes solutions, P5. L'enseignant comme soutien, P6. Briefing, P7. Débriefing, P8. Travailler en équipe multi point de vue et P9. Rapport d'analyse post-jeu. Ces patrons ont ensuite été présentés à 4 équipes, qui concevaient de nouveaux JENs avec le modèle LEGADEE. Nos observations montrent que la quasi-totalité des 14 concepteurs a trouvé ces patrons de conception très utiles et qu'ils ont amélioré leurs LGs en intégrant entre 1 à 5 de ces patrons.

Même si nos premiers résultats concernant l'utilisation du modèle LEGADEE pour analyser des JENs sont très prometteurs, ceux-ci ne représentent qu'une petite catégorie de LGs. De plus, le travail manuel d'analyse des scénarios, afin d'y trouver des invariants, est fastidieux et ne peut pas prétendre à être exhaustif. Une solution plus méthodique de trouver des patrons serait de proposer des outils d'analyse de scénario automatiques. Dans cette optique, plusieurs améliorations pourraient être apportées au modèle LEGADEE, notamment pour modéliser formellement la durée des étapes du *Scénario ludique*, les modalités des interactions entre les *Participants* (compétition, collaboration) et les ressorts de jeu mis en place tels que le hasard le vertige ou la reconnaissance [18]. En outre, un tel outil pourrait être utilisé afin d'analyser automatiquement le scénario d'un LG en cours de conception, afin de proposer aux concepteurs d'y intégrer des patrons qui n'y sont pas encore. Ceci semble correspondre à un réel besoin, puisque 10 sur les 14 concepteurs novices interrogés trouvent qu'un tel outil serait « utile » à « très utile ». Ce concept peut également être poussé plus loin, en proposant une coquille de scénario qui intègre un certain nombre de patrons, pour chaque type de LG. Par exemple, si l'on se base sur les patrons proposés dans cet article, le modèle initial d'un JEN pourrait contenir deux *Participant*s, un pour l'enseignant et l'autre pour un groupe de 3 à 4 élèves, ainsi que les *Personnages* associés (un personnage serviable et un groupe d'experts), ce qui correspond aux patrons P3 et P5. Le *Scénario ludique* peut également commencer par une *Mission teaser* (P1) et chaque nouvelle *Mission* créée sera automatique composée d'une *Séquence* briefing (P6) initiale et finira par une *Séquence* de débriefing (P7). Ces éléments initiaux serviraient de base aux concepteurs en panne d'inspiration et pourraient, bien évidemment, être modifiés ou effacés. Là encore, 10 sur les 14 concepteurs interrogés trouvent qu'un tel modèle initial serait « utile » à « très utile ».

## Remerciements

## Références

## Annexes

**P2. Problèmes pluridisciplinaires**
*Définition* : Le LG met en scène des problèmes qui font intervenir simultanément des compétences de plusieurs disciplines.
*Exemples* :
- LS : tous les *Modules* portent simultanément sur les compétences liées à l'urbanisme et le travail d'équipe ou la communication (Fig. 4, gauche).

- PU : tous les *Modules* portent sur les compétences liées au travail d'équipe et à la résolution de problème (Fig. 4, droite).

***Représentation avec le modèle LEGADEE :*** Tous les *Modules* doivent être liés à au moins deux *Compétences* de disciplines différentes.

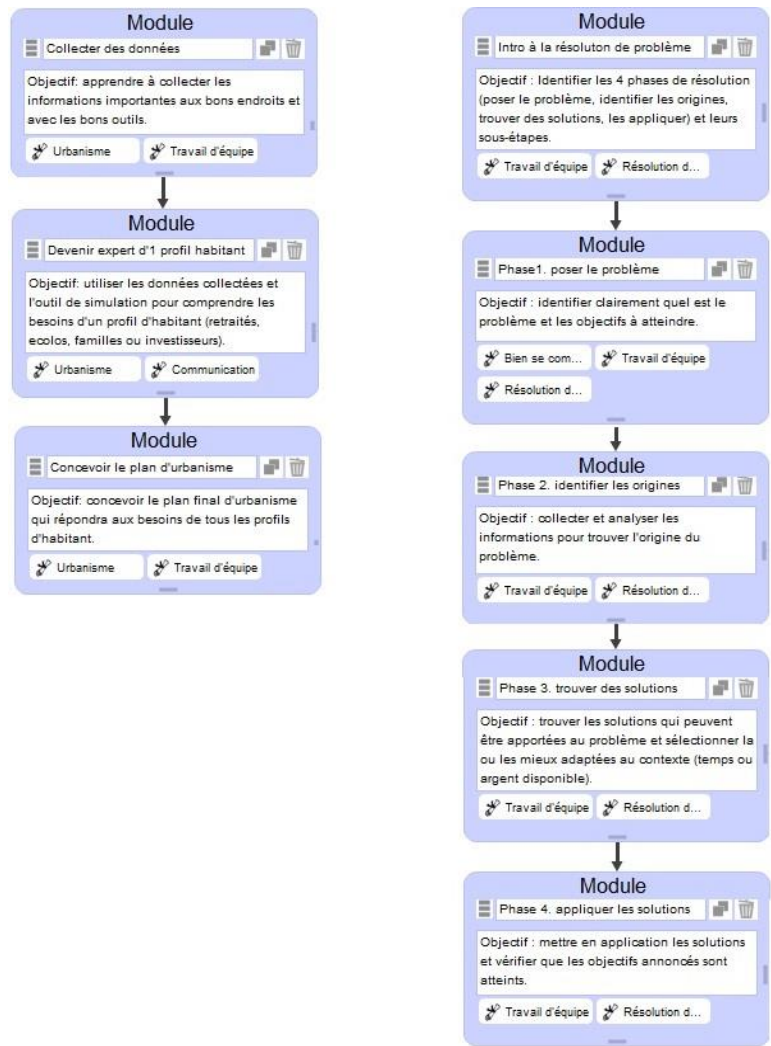

**Fig. 4.** Modélisation de P2. Problèmes pluridisciplinaires (LS-gauche, PU-droite)

**P3. Personnifier un groupe d'experts**
***Définition :*** Les apprenants sont mis par petit groupe (2 à 4 apprenants) et jouent le rôle d'experts du domaine.
***Exemples :***

- LS : les apprenants, par groupe de 3 à 4, jouent le rôle de stagiaires dans une entreprise d'urbanisme (Fig. 5, gauche).
- PU : le groupe de 2 à 3 apprenants joue le rôle d'une équipe de consultants, experts en résolution de problème (Fig. 5, droite).

**Représentation avec le modèle LEGADEE :**
- Les apprenants sont regroupés par équipe de *Participant* de 2 à 4.
- Cette équipe de *Participant* joue le rôle d'un *Personnage* du jeu qui représente des experts du domaine étudié.

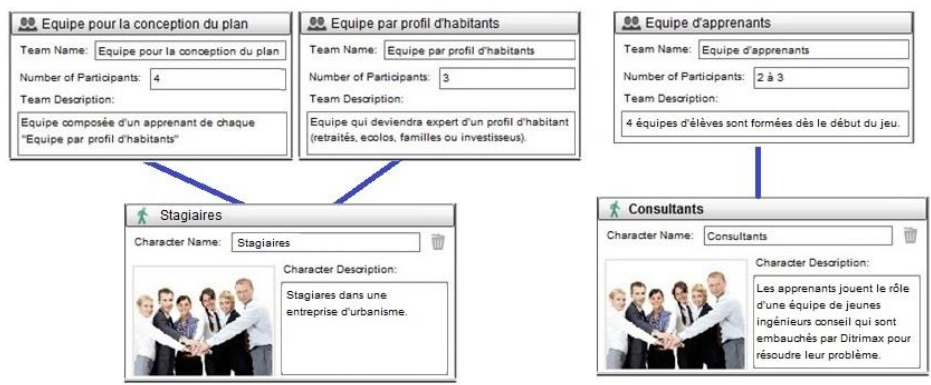

**Fig. 5.** Modélisation de P3. Personnifier un groupe d'expert (LS-gauche, PU-droite)

**P4. Explorer différents chemins**
*Définition* : Le scénario du jeu permet aux apprenants d'explorer plusieurs chemins afin de proposer des solutions différentes aux problèmes à résoudre.
*Exemples :*
- LS : Lors de la deuxième Mission du LGs, chaque groupe d'apprenants doit récolter des informations concernant les besoins d'un type d'habitant (retraités, écolos, familles et investisseurs) (Fig. 6, gauche).
- PU : Une fois que les apprenants ont défini les différents problèmes qui sont à l'origine des défaillances de l'entreprise, chaque groupe choisit le problème auquel il souhaite apporter une solution (Fig. 6, droite).

**Représentation avec le modèle LEGADEE :** *l*e *Scénario ludique* contient des embranchements parallèles.

**P6. Briefing**
*Définition* : L'enseignant, qui personnifie un tuteur ou un collègue bienveillant, met en place des sessions de briefing au début de chaque étape importante du LG, pour aider les joueurs à surmonter les défis.
*Exemples :*
- LS : chaque mission commence avec un briefing pendant laquelle les collègues serviables (joués par les enseignants) expliquent aux stagiaires (joués par les apprenants) les tâches qu'ils vont devoir accomplir et comment ils peuvent s'y préparer en organisant leur travail de groupe et en utilisant les outils pertinents. Avant de commencer les interviews avec les habitants par exemple, ils leur suggèrent d'écouter attentivement en prenant des notes sur les besoins exprimés par les habitants (Fig. 6, gauche).
- PU : chaque mission commence avec un briefing pendant lequel le superviseur (joué par l'enseignant) explique aux consultants (joués par les apprenants) en quoi consistent les tâches qu'ils devront accomplir et ce qu'elles vont leur apporter en terme de compétence et quel type de diagramme de décision ils peuvent utiliser pour les aider (Fig. 6, droite).

***Représentation avec le modèle LEGADEE :*** chaque *Mission* commence avec une *Séquence* de briefing. Cette *Séquence* est liée au *Personnage* de tuteur bienveillant, qui est joué par l'enseignant et au *Personnage* joué par les apprenants.

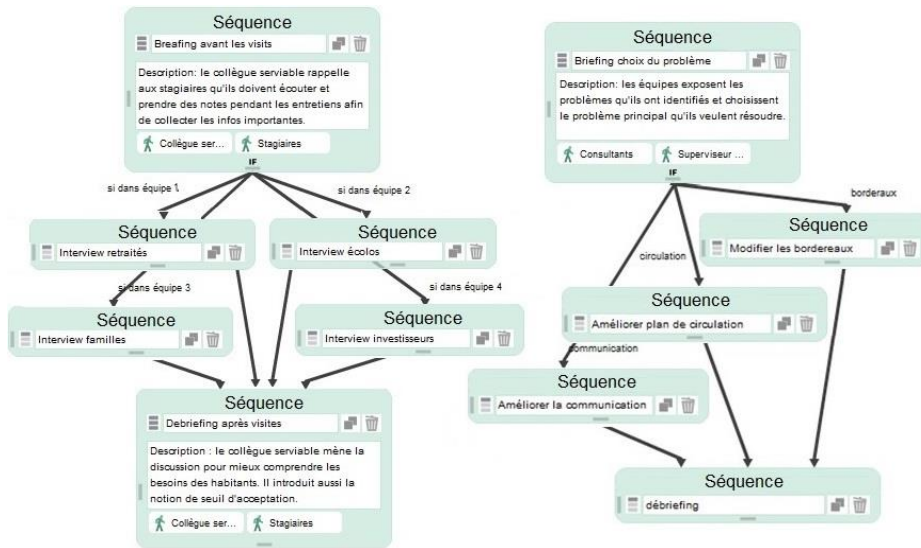

**Fig. 6.** Modélisation de P4. Explorer différents chemins (LS-gauche, PU-droite)

**P7. Débriefing**
***Définition :*** L'enseignant met en place des sessions de débriefing à la fin de chaque étape importante du jeu pour que les joueurs puissent partager leurs expériences, débattre des solutions trouver et comprendre comment ils peuvent réutiliser les compétences acquises dans d'autres contextes.
***Exemples :***
- LS : chaque mission finit par un débriefing avec tous les groupes de stagiaires. Pendant ces débriefings, les collègues serviables (joués par les enseignants) mettent aussi en avant les similitudes et les complémentarités des plans d'urbanisme proposés (Fig. 6, gauche).
- PU : Les quatre missions du jeu finissent par un débriefing pendant laquelle chaque groupe de consultant expose ses résultats et les méthodes employées pour y parvenir afin que tous puissent profiter de leurs expériences. Le superviseur (joué par l'enseignant) conclut en mettant en avant les compétences acquises et comment elles peuvent être utilisées dans d'autres situations (Fig. 6, droite).
***Représentation avec le modèle LEGADEE :*** Chaque *Mission* finit avec une *Séquence* de débriefing. Cette *Séquence* est liée au *Personnage* de tuteur bienveillant, qui est joué par l'enseignant et au *Personnage* joué par les apprenants. Afin d'enrichir le débat et les retours d'expérience, les joueurs s'appuient sur les *Documents* qu'ils ont trouvés ou écrits pendant le jeu.

### P8. Travailler en équipe multi point de vue
***Définition :*** Afin de favoriser des débats, les apprenants jouent une partie du jeu en équipe composée de personnes ayant des points de vue divergents.
***Exemple :***
- LS : la première partie du LG consiste à concevoir un plan d'urbanisme qui convient à un type d'habitant (retraités, écolos, familles et investisseurs). À l'issue de cette étape, qui implique de nombreux échanges avec les habitants concernés, les stagiaires maîtrisent parfaitement leurs besoins. Pour la suite du LG, les équipes sont recomposées de façon à avoir un représentant de chaque type d'habitants. Cette composition permet ainsi de concevoir le plan d'urbanisme final qui doit prendre en compte les besoins de tous les habitants. (Fig. 6, gauche).

***Représentation dans le modèle LEGADEE :*** À un moment donné dans le LG, les groupes de *Participants* sont composés de plusieurs apprenants qui ont des points de vue divergents.

### P9. Rapport d'analyse post-jeu
***Définition :*** Les apprenants ne sont pas évalués sur leurs actions pendant le jeu, mais sur un rapport d'analyse à rédiger après le jeu.
***Exemple :***
- PU : Les apprenants sont évalués uniquement sur le rapport d'analyse qu'ils doivent rendre une semaine après le LG. Ce rapport doit expliciter les compétences acquises pendant le LG et être enrichi de recherches personnelles sur les méthodes de résolutions de problème (Fig. 7).

***Représentation dans le modèle LEGADEE :*** la dernière *Mission* du jeu consiste à écrire un rapport sur les compétences acquises pendant le jeu.

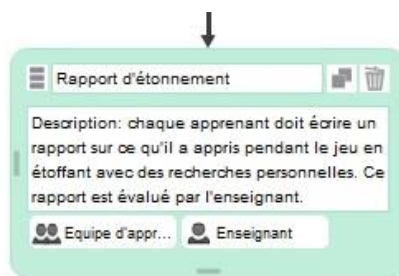

**Fig. 7.** Modélisation de P9. Rapport d'analyse post-jeu